\documentclass[showpacs,preprintnumbers,amsmath,amssymb]{revtex4}
\usepackage{graphicx}
\usepackage{dcolumn}
\usepackage{bm}

\newcommand{\rmd}{{\rm d}}

\newcommand{\tr}{{\text{Tr}}}

\begin{document}

\title{Frequency-dependent current correlation functions from scattering
  theory}

\author{J. Salo}

\affiliation{Laboratory of Physics, Helsinki University of
  Technology, PO Box 4100, 02015 TKK, Finland}

\author{F.W.J. Hekking}

\affiliation{Laboratoire de Physique et Mod\'elisation des Milieux
Condens\'es, CNRS \& Universit\'e Joseph Fourier, BP 166, 38042
Grenoble cedex 09, France }

\author{J.P. Pekola}

\affiliation{Low Temperature Laboratory, Helsinki University of
  Technology, PO Box 3500, 02015 TKK, Finland}

\begin{abstract}
  We present a general formalism based on scattering theory to calculate
  quantum correlation functions involving several time-dependent current
  operators. A key ingredient is the causality of the scattering matrix, which
  allows one to deal with arbitrary correlation functions. Our
  formalism might be useful in view of recent developments in full counting
  statistics of charge transfer, where detecting schemes have been proposed
  for measurement of frequency dependent spectra of higher moments.  Some of
  these schemes are different from the well-known fictitious spin-detector and
  therefore generally involve calculation of non-Keldysh-contour-ordered
  correlation functions. As an illustration of our method we consider
  various third order correlation functions of current, including the usual third
  cumulant of current statistics. We investigate the frequency dependence of these
  correlation functions explicitly in the case of energy-independent scattering.
  The results can easily be generalized to the calculation of arbitrary $n$-th order
  correlation functions, or to include the effect of interactions.
\end{abstract}


\maketitle

\section{Introduction}

Dynamical noise properties of mesoscopic systems have been studied
for more than a decade, both theoretically and
experimentally~\cite{blanter00}. By now it is well understood that
noise measurements can reveal information on the system that is not
contained in its DC conductance. So far, most experiments
concentrated on measurement of zero-frequency noise. However,
several proposals have considered the possibility of detecting
finite-frequency noise, for instance through emission and absorption
measurements using quantum few level systems like quantum
dots~\cite{aguado00} or small Josephson
junctions~\cite{schoelkopf03} as noise detectors. Successful
experiments of this type have been reported
recently~\cite{deblock0306,astafiev04}. Finite frequency noise is
interesting, first of all as one expects the noise to probe the
intrinsic dynamics of the conductor and hence the noise spectral
function should be sensitive to the dwell time $\tau _D$ of the
carriers. Second, at finite frequency, current is no longer
spatially homogeneous, and charge piles up in the conductor. Coulomb
interaction screens this pile-up of charge, at a characteristic
charge relaxation frequency $1/\tau$ which may well be different
from $1/\tau_D$. These issues have been studied theoretically for
diffusive contacts in Refs.~\cite{naveh97,nagaev98}. Recent
calculations of current noise in chaotic
cavities~\cite{nagaev04,hekking05} that take both the
energy-dependence of scattering and Coulomb interactions into
account show that the frequency-dependent noise spectrum is
determined solely by the time $\tau$, as long as quantum corrections
like weak-localization can be ignored. In view of recent interest in
the theory of the full counting statistics (FCS) of charge
transfer~\cite{nazarov03}, attention shifted from the conventional
noise to the study of the properties of the higher moments. Recent
measurements have probed the zero-frequency third
cumulant~\cite{reulet03,bomze05,gustavsson:076605}. As far as the frequency dependence
of the higher cumulants is concerned, the situation changes
drastically as compared to conventional noise spectra. Calculations
of the frequency-dependent third cumulant for a chaotic
cavity~\cite{nagaev04} and for a diffusive
conductor~\cite{pilgram04} show marked differences from the
conventional noise: it is not only determined by the
charge-relaxation time $\tau$ but also shows a low-frequency
dispersion that is determined by the dwell time $\tau_D$.

A properly designed experiment, capable of measuring the frequency-dependent
third cumulant, would thus enable one to determine the two relevant time scales
separately in a mesoscopic conductor. The question as to how to design such an
experiment brings us to one of the key problems of this field: what is an
adequate detector to measure frequency-dependent noise spectra, and which
noise spectral function is it actually measuring? Most of the applications of
FCS discussed so far concentrate on the use of a fictitious spin detector,
introduced by Levitov and coworkers~\cite{levitov94,levitov96}. This detector
measures Keldysh contour-ordered correlation functions of current. Powerful
theoretical tools have been developed to calculate these correlation
functions; therefore this detector is amenable to straightforward analysis.
However, the spin detector might not be the most suitable one for detecting
finite frequency noise. Detectors that interact with the noise source through
emission and absorption, like the abovementioned quantum detectors might be
more suitable for this task. The measured spectra are then not directly
related to Keldysh-ordered correlation functions, and different methods are
required to determine these spectra theoretically.

In this paper we develop a method capable of handling arbitrarily ordered
correlation functions. The formalism we adopt is based on scattering
theory~\cite{buttiker92}, pioneered in~\cite{lesovik89,yurke90,buttiker90}. It is
the natural approach to discuss transport and noise in mesoscopic
devices. The operator for electric
current $\hat{I}$ is written as the difference between the current
carried by incident particles $\hat{I}_\text{in}$ and the current
carried by scattered particles $\hat{I}_\text{out}$: $\hat{I} =
\hat{I}_\text{in} - \hat{I}_\text{out}$. The central quantity of the
scattering approach is the energy-dependent scattering matrix.  It
must satisfy the causality condition in real-time representation,
which has immediate consequences for the commutation relations
between the operators $\hat{I}_\text{in}$ and $\hat{I}_\text{out}$
at different times~\cite{beenakker01}. As a result, {\em any} (anti)
time-ordered product of current operators can be conveniently
rewritten as products of currents $\hat{I}_\text{in}$ and
$\hat{I}_\text{out}$ with all in-currents ordered to the right
(left) of the out-currents. Denoting (anti) time-ordering by $T$
($\tilde{T}$) this implies $T[\hat{I}_\text{in}(t_1)
\hat{I}_\text{out}(t_2)] = \hat{I}_\text{out}(t_2)
\hat{I}_\text{in}(t_1)$ and $\tilde{T}[\hat{I}_\text{in}(t_1)
\hat{I}_\text{out}(t_2)] = \hat{I}_\text{in}(t_1)
\hat{I}_\text{out}(t_2)$ {\em independent of the ordering of $t_1$
and $t_2$}. This way, the cumbersome time-ordering can be avoided
and the remaining in-out~-ordered products can be readily calculated
using the scattering theory.

We apply the in-out~-ordering method to the well-studied case of the
third cumulant of charge transfer in a mesoscopic conductor. We
treat energy-independent scattering, and present the time-dependent
cumulant in the cases of a tunnel barrier (a quantum point contact),
a diffusive wire, and a chaotic cavity. First of all, this enables a
direct check on the validity of our method. Second, we believe that
the zero frequency limit of the calculation provides a demonstration
of the validity of the result for the third cumulant of a tunnel
barrier presented in~\cite{levitov04}. This result had given rise to
some discussion in the
literature~\cite{levitov92,levitov94,levitov96} and methods have
been developed to settle the issue in a
frequency-dependent context~\cite{lesovik03,galaktionov03}. Thirdly,
our calculation of the frequency-dependent third cumulant enables us
to find the asymptotic time-dependence of the third cumulant of the
charge transfer, both in the short and the long time limits.

The paper is organized as follows: we first summarize the scattering
formalism in order to define the notation used later, and use the
causality of the scattering matrix to derive important commutation
relations between in- and out-current operators. They are used to
establish operator transformation rules, such as
$T[\hat{I}_\text{in}(t_1) \hat{I}_\text{out}(t_2)] =
\hat{I}_\text{out}(t_2) \hat{I}_\text{in}(t_1)$, which allow one to
resolve time-ordered products of currents in terms of
in-out~-ordered products. Their main application is to find
multi-current correlation functions, and we explicitly present all
three-current correlations, which are written in terms of
three-current spectral functions of two frequency arguments. To keep
the presentation transparent, we do not address here issues
concerning the finite dwell time of carriers nor do we address
interaction effects. We thus treat the case of energy-independent
scattering where the various spectral functions can be evaluated
using only the transmission probabilities of the scatterer, valid in
the limit where the above-mentioned characteristic times $\tau _D$,
$\tau$ vanish. It is important to note that, even though we neglect
the energy dependence of the scattering matrix, we do respect its
causality through the in-out~-ordering properties. We finally
discuss several different detection schemes, which all correspond to
different three-current correlation functions and, most importantly,
use the full-counting statistics approach to derive an expression to
the time-dependent third cumulant of transmitted charge
distribution.

\section{Scattering formalism and causality}

\subsection{Scattering theory}

\begin{figure}
    \begin{center}
    \includegraphics[width=8.0cm]{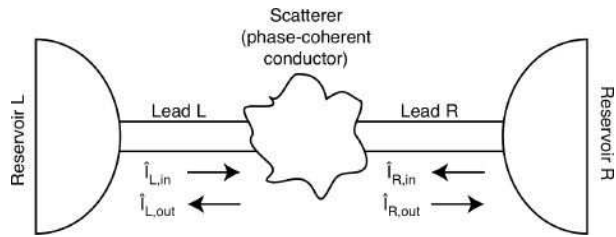}
    \end{center}
    \caption{Two-terminal scattering problem. Both reservoirs are assumed to be in thermal equilibrium, characterized by a common temperature $T$ and potentials such that $V_R-V_L=V$.}
    \label{F1}
\end{figure}

The starting point for our analysis is scattering theory, as developed by
B\"uttiker~\cite{buttiker92}. In this formalism, the current operator of
non-interacting electrons is given by
\begin{equation}
\begin{split}
\hat I_\alpha (t) &= \frac{e}{h} \sum_n \int dE dE'
e^{i(E-E')t/\hbar}
  [ \hat a^{\dagger}_{\alpha n} (E) \hat a_{\alpha n} (E')
        -\hat b^{\dagger}_{\alpha n} (E) \hat b_{\alpha n} (E')].
\label{curalpha}
\end{split}
\end{equation}
The operators $\hat a^{\dagger}_{\alpha n} (E)$ and $\hat a_{\alpha n} (E)$
create and annihilate electrons with total energy $E$ in the transverse
channel $n$ in lead $\alpha$, incident upon the scatterer. Similarly, the
creation $\hat b^{\dagger}_{\alpha n} (E)$ and annihilation $\hat b_{\alpha n}
(E)$ operators refer to electrons in the outgoing states. For the two-terminal
set-up depicted in Fig.~\ref{F1}, $\alpha$ takes values $L$ and $R$ for the
left and right leads respectively. The results to be presented below can be
easily generalized to any multi-terminal case.
The creation and annihilation operators obey the anticommutation relations, for
instance,
\begin{equation} \label{comm}
\begin{split}
\hat a^{\dagger}_{\alpha n}(E) \hat a_{\alpha n'}(E')
  +\hat a_{\alpha n'}(E')\hat a^{\dagger}_{\alpha n}(E)
&= \delta_{nn'} \delta(E - E')\\
\hat a^{\dagger}_{\alpha n}(E)\hat a^{\dagger}_{\alpha n'}(E')
  +\hat a^{\dagger}_{\alpha n'}(E') \hat a^{\dagger}_{\alpha n}(E)
&= 0\\
\hat a_{\alpha n}(E) \hat a_{\alpha n'}(E')
  +\hat a_{\alpha n'}(E') \hat a_{\alpha n}(E)
&= 0.
\end{split}
\end{equation}
Similar anticommutation relations hold naturally also for operators referring
to the outgoing states.

The operators $\hat a$ and $\hat b$ are related by the scattering matrix $s$,
\begin{equation} \label{bvsa}
\hat b_{\alpha n}(E)
= \sum_{\beta,m}s_{\alpha \beta;nm}(E) \hat{a}_{\beta m}(E)
\end{equation}
and the creation operators $\hat a^{\dagger}$ and $\hat b^{\dagger}$
are correspondingly related by the hermitian conjugated matrix,
$s^{\dagger}_{\alpha \beta;nm}(E) = s^{*}_{\alpha \beta;mn}(E)$.

The matrix $s$ is quite generally unitary and it has dimensions $(N_L + N_R)
\times (N_L + N_R)$. Its size and the matrix elements depend on the total
energy $E$. It has the block structure
\begin{equation} \label{smatr}
s = \left(\begin{array}{cc}
    r & t' \\
    t & r'
    \end{array}\right).
\end{equation}
Electron reflection back to the left and right reservoirs is described by the
square diagonal blocks $r$ (size $N_L \times N_L$) and $r'$ (size $N_R \times
N_R$), respectively, while the off-diagonal, rectangular blocks $t$ (size $N_R
\times N_L$) and $t'$ (size $N_L \times N_R$) determine, in turn, the electron
transmission through the sample.

In order to directly benefit from consequences of causality, we
present the current operator as the difference between two {\em
directed} currents, carried by incoming states and outgoing states,
respectively~\cite{beenakker01}.  Specializing to the left lead, we
thus write
\begin{equation}
\hat I_L (t)  =  \hat I_{L,\text{in}} (t) - \hat I_{L,\text{out}} (t),
\end{equation}
where
\begin{equation}
\hat I_{L,\text{in}} (t) =
 \frac{e}{h} \sum_n
\int dE dE' e^{i(E-E')t/\hbar} \hat a^{\dagger}_{Ln} (E) \hat a_{Ln} (E') ,
\end{equation}
and
\begin{equation}
\hat I_{L,\text{out}} (t) =
 \frac{e}{h} \sum_n
\int dE dE' e^{i(E-E')t/\hbar} \hat b^{\dagger}_{Ln} (E) \hat b_{Ln} (E') .
\end{equation}
Now, using Eq. (\ref{bvsa}) as well as its hermitian conjugated version, $\hat
I_{L,\text{out}} (t)$ can also be written as
\begin{equation}
\begin{split}
\hat I_{L,\text{out}} (t) = \frac{e}{h} \sum_{\alpha,\beta} \sum_{mnk}
   \int dE dE' e^{i(E-E')t/\hbar}
   \hat a^{\dagger}_{\alpha m}(E)
   s^{\dagger}_{L\alpha;mk}(E) s_{L\beta;kn}(E')\hat a_{\beta n} (E')
   ,
\label{outcur}
\end{split}
\end{equation}
where indices $\alpha$ and $\beta$ may take values $L$ or $R$. This
result makes the dependence of the current operator
$\hat{I}_\mathrm{out}$ on the energy-dependent scattering matrix
$s(E)$ explicit. As we will detail below, the commutation properties
of directed current operators at different times are completely
determined by the analytical properties of $s(E)$.

\subsection{Causality}

In real time, the scattering matrix connects operators of an
incoming state with those of an outgoing state by the convolution
relation
\begin{equation}
\hat b_{\alpha n}(t)
= \sum_{\beta,m} \int_{-\infty}^\infty
  s_{\alpha \beta;nm}(t-\tau) \hat{a}_{\beta m}(\tau) \rmd\tau.
\end{equation}
By causality, the scattering matrix must vanish for negative
arguments since otherwise an incident current at $t_1$ would cause
an outgoing current at $t_2<t_1$. This is equivalent to requiring that
the Fourier transform of the scattering matrix, $s_{\alpha\beta;nm}(\omega)$
be analytic in the entire upper half plane, since then
\begin{equation}
s_{\alpha \beta;nm}(\omega)
=\lim_{\eta\rightarrow 0^+} \int_{-\infty}^{\infty}\frac{d\omega'}{2\pi i}
 \frac{s_{\alpha\beta;nm}(\omega')}{(\omega'-\omega)-i\eta},
\end{equation}
which can be substituted into the inverse Fourier transform of the
scattering matrix in order to obtain
\begin{equation}
\begin{split}
s_{\alpha\beta;nm}(t)
&= \int_{-\infty}^{\infty}\frac{d\omega'}{2\pi}
   s_{\alpha\beta;nm}(\omega')
   \int_{-\infty}^{\infty}\frac{d\omega}{2\pi i}e^{-i\omega t}
   \frac{1}{(\omega'-\omega)-i\eta}
 = \theta(t) s_{\alpha\beta;nm}(t).
\end{split}
\label{theta}
\end{equation}
Hence the analytical structure of $s(\omega)$ as a function of
$\omega$ (analyticity in the entire upper half plane) implies
causality~\cite{shepard91,buttiker93}, i.e., $s(t-t') = 0$ if
$t<t'$. Similarly, the hermitian conjugated scattering matrix,
$s^\dagger_{\alpha \beta;nm}(\omega)$, must be analytic in the
entire lower half plane.

\subsection{Commutation relations}\label{comrel}

We will use the analytical structure of the scattering matrix
established in the previous subsection, Eq. \eqref{theta}, to obtain
the commutation relations for directed current operators
$\hat{I}_\mathrm{in}$ and $\hat{I}_\mathrm{out}$ at different
times~\cite{beenakker01}. Consider the commutation relation of $\hat
I_{L,\text{in}} (t_1)$ and $\hat I_{L,\text{out}} (t_2)$. Starting
from
\begin{equation}
\begin{split}
&[\hat I_{L,\text{in}} (t_1),\hat I_{L,\text{out}} (t_2)]\\
&= \left(\frac{e}{h}\right)^2
   \sum \limits _{n_1,n_2}  \int dE_1 dE_2 dE_3 dE_4 e^{i(E_1-E_2)t_1/\hbar}
e^{i(E_3-E_4)t_2/\hbar}  [\hat a^{\dagger}_{Ln_1} (E_1) \hat
a_{Ln_1} (E_2) , \hat b^{\dagger}_{Ln_2} (E_3) \hat b_{Ln_2} (E_4)],
\end{split}
\end{equation}
and applying the commutation relations as given in \eqref{comm} we
find that
\begin{equation}
\begin{split}
& [\hat I_{L,\text{in}} (t_1),\hat I_{L,\text{out}} (t_2)]\\
&= \left(\frac{e}{h}\right)^2
   \sum_{n_1,n_2}  \int dE_1 dE_2 dE_3
   \big[ e^{i(E_1-E_3)t_1/\hbar}
         e^{i(E_3-E_2)t_2/\hbar}
         \hat{a}^{\dagger}_{Ln_1}(E_1)
         s^{\dagger}_{LL;n_1n_2}(E_3)
         \hat{b}_{Ln_2}(E_2)\\
&\qquad\qquad\qquad\qquad\qquad\qquad\qquad
        -e^{i(E_3-E_2)t_1/\hbar}
         e^{i(E_1-E_3)t_2/\hbar}
         \hat{b}^{\dagger}_{Ln_2} (E_1)
         s_{LL;n_2 n_1}(E_3)
         \hat{a}_{Ln_1}(E_2)
    \big].
\end{split}
\end{equation}
Integrating over all energies we obtain
\begin{equation} \label{inoutc}
\begin{split}
[\hat I_{L,\text{in}} (t_1),\hat I_{L,\text{out}} (t_2)]
= h e^2 \sum_{n_1,n_2}
        \hat{a}^{\dagger}_{Ln_1}(t_1)
        s^{\dagger}_{LL;n_1n_2}(t_2-t_1)
        \hat{b}_{Ln_2}(t_2)
       -\hat{b}^{\dagger}_{Ln_2} (t_2)
        s_{LL;n_2n_1}(t_2-t_1)
        \hat{a}_{Ln_1}(t_1).
\end{split}
\end{equation}
According to Eq. \eqref{theta} the commutator \eqref{inoutc}
vanishes identically if $t_1$ is a later instant of time than
$t_2$~\cite{beenakker01}. We thus conclude that
\begin{equation} \label{commutator1}
[\hat I_{L,\text{in}} (t_1),\hat I_{L,\text{out}} (t_2)] \propto
\theta(t_2-t_1).
\end{equation}

We obtain the commutation relations for $\hat I_{L,\text{in}} (t_1)$
and $\hat I_{L,\text{in}} (t_2)$, and for $\hat I_{L,\text{out}}
(t_1)$ and $\hat I_{L,\text{out}} (t_2)$ using the same procedure:
both these vanish identically,
\begin{equation} \label{commutator2}
[\hat I_{L,\text{in}} (t_1),\hat I_{L,\text{in}} (t_2)] = 0
\end{equation}
and
\begin{equation} \label{commutator2b}
[\hat I_{L,\text{out}} (t_1),\hat
I_{L,\text{out}} (t_2)] = 0.
\end{equation}
These commutation relations have important consequences for the
calculation of time-ordered correlation functions involving the
operators $\hat{I}_\text{in}(t)$ and $\hat{I}_\text{out}(t)$, as we
will now show.

\subsection{Time-ordered correlation functions}
We denote the time-ordering of operators by $T[A(t_1)B(t_2)C(t_3)\dots]$,
where the operators appear in descending order of times, and the
anti-time-ordering by $\tilde{T}[A(t_1)B(t_2)C(t_3)\dots]$, with the
opposite order of times. Specifically, making use of (\ref{commutator1}),
(\ref{commutator2}), and (\ref{commutator2b}), we find the following
operator identities:
\begin{equation}
\begin{split}
T[\hat{I}_\text{in}(t_1) \hat{I}_\text{in}(t_2)]
&= \hat{I}_\text{in}(t_1) \hat{I}_\text{in}(t_2), \\
T[\hat{I}_\text{out}(t_1) \hat{I}_\text{out}(t_2)]
&= \hat{I}_\text{out}(t_1) \hat{I}_\text{out}(t_2), \\
T[\hat{I}_\text{in}(t_1) \hat{I}_\text{out}(t_2)]
&= \hat{I}_\text{out}(t_2) \hat{I}_\text{in}(t_1), \\
T[\hat{I}_\text{out}(t_1) \hat{I}_\text{in}(t_2)]
&= \hat{I}_\text{out}(t_1) \hat{I}_\text{in}(t_2).
\end{split}
\end{equation}
One therefore concludes~\cite{beenakker01}: {\em time-ordering a
product of directed current operators corresponds to an ordering in
which all the out-currents $\hat{I}_\mathrm{out}$ are placed to the
left of the in-currents $\hat{I}_\mathrm{in}$.}

As an example, let us consider the two lowest time-ordered
correlation functions. Using $\hat{I}(t) = \hat{I}_\text{in}(t) -
\hat{I}_\text{out}(t)$, one obtains
\begin{equation}
\begin{split}
&T[\hat{I}(t_1) \hat{I}(t_2)]= \hat{I}_\text{in}(t_1)
\hat{I}_\text{in}(t_2)
  -\hat{I}_\text{out}(t_2) \hat{I}_\text{in}(t_1)
  -\hat{I}_\text{out}(t_1) \hat{I}_\text{in}(t_2)
  +\hat{I}_\text{out}(t_1) \hat{I}_\text{out}(t_2),
\label{eq:two-current}
\end{split}
\end{equation}
and
\begin{equation}
\begin{split}
T[\hat{I}(t_1) \hat{I}(t_2)\hat{I}(t_3)]
&= \hat{I}_\text{in}(t_1)\hat{I}_\text{in}(t_2)\hat{I}_\text{in}(t_3)
  -\hat{I}_\text{out}(t_3) \hat{I}_\text{in}(t_1) \hat{I}_\text{in}(t_2)
  -\hat{I}_\text{out}(t_2) \hat{I}_\text{in}(t_1) \hat{I}_\text{in}(t_3)\\
&\quad
  -\hat{I}_\text{out}(t_1) \hat{I}_\text{in}(t_2) \hat{I}_\text{in}(t_3)
  +\hat{I}_\text{out}(t_2) \hat{I}_\text{out}(t_3) \hat{I}_\text{in}(t_1)
  +\hat{I}_\text{out}(t_1) \hat{I}_\text{out}(t_3) \hat{I}_\text{in}(t_2)\\
&\quad
  +\hat{I}_\text{out}(t_1) \hat{I}_\text{out}(t_2) \hat{I}_\text{in}(t_3)
  -\hat{I}_\text{out}(t_1) \hat{I}_\text{out}(t_2)\hat{I}_\text{out}(t_3).
\end{split}
\label{eq:three-current}
\end{equation}
For the ordered $n$-current correlation function, the number of
terms containing $p$ out-currents and $n-p$ in-currents is just the
binomial factor $n!/[p!(n-p)!]$. The sign of such a term is
$(-1)^p$. The anti-time-ordering $\tilde{T}$ can be dealt with analogously,
but here the in and out currents are ordered oppositely: all the
out-currents stand to the {\em right} of the in-currents.

The important point here, and one of the central conclusions of
Ref.~\cite{beenakker01}, is that using in-out ordering one gets rid
of the cumbersome limits of time integration, normally present in
time-ordered expressions. This will enable us in the following to
straightforwardly calculate Fourier transforms and hence directly
obtain the frequency-dependent spectral functions of the relevant
correlation functions. Moreover, the idea of ordering currents using
the in-out formalism is quite natural in scattering theory.

\section{In-out three-current spectral functions}

\subsection{General results}
\begin{figure*}
\begin{center}
\includegraphics[width=16cm]{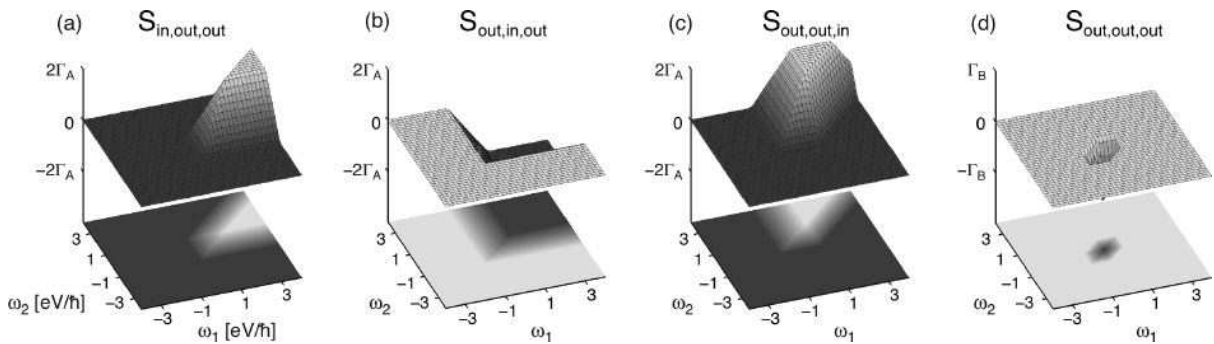}
\end{center}
\caption{\label{fig:noises}
  The non-vanishing contributions to zero-temperature three-current spectral functions
  $S_\text{in,out,out}$, $S_\text{out,in,out}$, $S_\text{out,out,in}$, and
  $S_\text{out,out,out}$ plotted against the frequencies $\omega_1$ and $\omega_2$.
  The first three of these have values between $0$ and $\pm 2\Gamma_\text{A} = \pm
  2eV\frac{e^3}{h}\sum_n T_n(1-T_n)$ while the last one has the extreme
  value of $-\Gamma_\text{B} = -eV \frac{e^3}{h} \sum_n T_n(1-T_n)(1-2T_n)$ at the origin
  and it vanishes at high frequencies, $|\omega_1|,|\omega_2|\gg eV/\hbar$.}
\end{figure*}

We now turn to consider various three-current correlation functions
of the form $\langle \delta \hat{I}_{L,x}(t_1) \delta
\hat{I}_{L,y}(t_2) \delta \hat{I}_{L,z}(t_3) \rangle$, where each of
the $x$, $y$, and $z$ refers to the directed component of the
current, either in or out, and $\delta \hat{I}=\hat{I} -
\langle\hat{I}\rangle$. In the time-independent case, they can be
expressed using the Fourier transform given by
\begin{equation}
\begin{split}
&\langle \delta \hat{I}_{L,x}(t_1)
               \delta \hat{I}_{L,y}(t_2)
               \delta \hat{I}_{L,z}(t_3)\rangle
               \equiv \int \frac{d\omega_1}{2 \pi}e^{- i\omega_1(t_1-t_2)}\int
\frac{d\omega_2}{2\pi}
 e^{- i\omega_2(t_2-t_3)} S_{xyz}(\omega_1, \omega_2),
\end{split}
\label{eq:Ftrans}
\end{equation}
where $S_{xyz}(\omega_1, \omega_2)$ are the corresponding
three-current spectral functions. (Note that another convention is
to take the transform with respect to $t_1-t_2$ and $t_1-t_3$, which
leads to slightly redefined parameterization of the spectral
functions.) Specializing to the case of equilibrium reservoirs, the
spectral functions $S_{xyz}(\omega_1, \omega_2)$ are obtained by
applying Wick's theorem; we refer the reader to
Appendix~\ref{app:Wick} for details. Specifically, we present
results for the three-current spectral functions in the general case
of an arbitrary energy-dependent scattering matrix in Table
\ref{tab:energydependent} of Appendix \ref{app:spectral}, and for
energy-independent scattering in Table \ref{tab:energyindependent}
in the same Appendix. Here we just note that for the particular case
of $S_\text{in,in,in}$, the energy integral contains Fermi functions
of only one reservoir, and its value vanishes then identically. This
is due to the fact that the {\em in-in-in} term does not contain the
possibly energy-dependent scattering matrix. Spectral functions
containing two in-currents also only depend on the Fermi function of
the left reservoir, but the energy-dependence of the scattering
matrix may render the integrals nonzero. Such terms, however, vanish
in the case of energy-independent scattering so that four spectral
functions out of the eight are identically zero. The four remaining
spectra at zero temperature are depicted in Fig.~\ref{fig:noises}
as functions of the two frequencies $\omega_1$ and $\omega_2$.

\subsection{Limiting cases of in-out~-ordered spectral functions}
Although the true advantage of in-out ordering comes when dealing
with general correlation functions, we demonstrate here that it also
provides a straightforward way to obtain the spectral functions in
some special cases which have been discussed in literature already
earlier. In particular, we investigate here the case of
energy-independent scattering in the limiting cases in terms of
temperature, voltage, and the two frequencies.

As mentioned above, in the case of energy-independent scattering
only four three-current spectral functions out of eight possible
ones remain nonzero. At zero frequencies, $\omega_1 = \omega_2 = 0$,
only $S_\text{out,in,out}$ and $S_\text{out,out,out}$ are finite, with
their values given by
\begin{equation}
\begin{split}
S_\text{out,in,out}
&=-eV \frac{e^3}{h} \sum_n T_n(1-T_n)
 \equiv -\Gamma_\text{A} \\
S_\text{out,out,out}
&=-eV \frac{e^3}{h} \sum_n T_n(1-T_n)(1-2T_n)
 \equiv -\Gamma_\text{B}
\end{split}
\end{equation}
where $\Gamma_\text{A}=e^2 GV F_2$ and $\Gamma_\text{B}=e^2 GV F_3$
are expressed in terms of the conductance, $G=\frac{e^2}{h}\sum_n
T_n$, and the Fano factors of the second and third order,
$F_2=\sum_n T_n(1-T_n)/\sum_n T_n$ and $F_3=\sum_n
T_n(1-T_n)(1-2T_n)/\sum_n T_n$. The transmission eigenvalues
$\{T_n\}$ are the eigenvalues of the transmission matrix $t^\dagger
t$. In the high-frequency limit, $|\omega_1|,|\omega_2|\gg
eV/\hbar$, the non-vanishing terms are in turn $S_\text{in,out,out}$
and $S_\text{out,out,in}$, whose values equal $2\Gamma_\text{A}$ in
the second and first octants, respectively, and
$S_\text{out,in,out}=-2\Gamma_\text{A}$ in the first quadrant.

At finite temperatures such that $\hbar |\omega_1|, \hbar |\omega_2|
\ll k_BT$, the spectral functions
become independent of $\omega_1$ and $\omega_2$, and the
non-vanishing ones are given by
\begin{equation}
\begin{split}
S_\text{in,out,out}
&= S_\text{out,out,in}
 = \frac{1}{3}\Gamma_\text{A}\\
S_\text{out,in,out}
&=-\frac{2}{3}\Gamma_\text{A}.
\end{split}
\end{equation}

\section{Different physical detector schemes}

An arbitrary three-current correlation function can always be
decomposed into a sum of various in-out~-ordered spectral functions
of the type of Eq.~(\ref{eq:Ftrans}), whose properties are, at least
in principle, known. We will illustrate the usefulness of this
decomposition scheme now for various examples of three-current
correlation functions which have appeared in the literature. For
simplicity we assume energy-independent scattering such that
definite results can be obtained for three specific examples. We
will first consider accumulated charge by a fictitious
spin-detector~\cite{levitov96}, which directly depends on the
Keldysh-ordered correlation functions, and we use the in-out
three-current spectral functions to evaluate time-dependent third
cumulant of the charge distribution. We also compare this with current
statistics derived from an unordered generating function and relate it to
some of the results earlier appeared in literature. The second example is a
classical detector which would correspond to the standard fully
symmetrized three-current correlation function~\cite{golubev05}, and
finally we briefly discuss a partially time-ordered correlation
function that appears when the time evolution of the density matrix
of a multilevel quantum detector is considered, coupled to a
non-gaussian noise source~\cite{ojanen06,brosco06}.

\subsection{Third cumulant of FCS}

\begin{figure}
    \begin{center}
    \includegraphics[width=16cm]{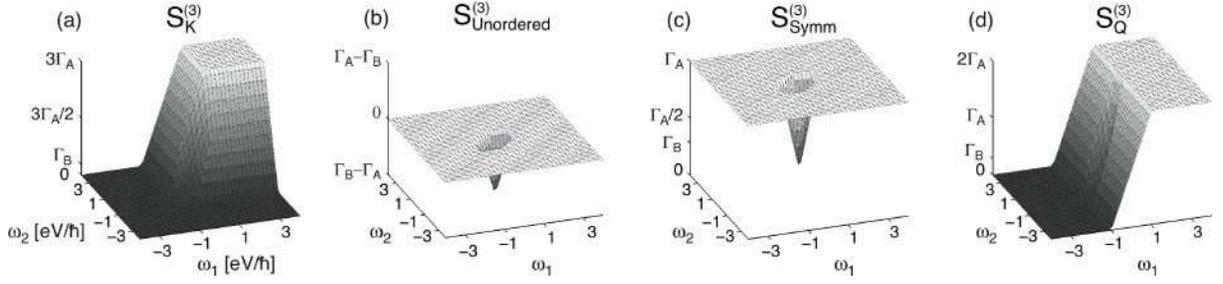}
    \end{center}
    \caption{(a) Keldysh-ordered, (b) unordered, (c) symmetrized, and
    (d) non-symmetrized spectra at zero temperature. The saturated levels
    of the spectral functions are all proportional to $\Gamma_\text{A}$
    except for the unordered spectrum, which  saturates to zero. For the
    other spectra the level essentially depends on in which areas of the
    $(\omega_1,\omega_2)$--plane the spectrum has non-zero values. Both
    $S_\text{in,out,out}$ and $S_\text{out,out,in}$ vanish at zero frequency,
    and the spectral functions (a), (c), and (d) are then determined by
    $\Gamma_\text{B}=-S_\text{out,out,out}$, while the spectrum (b) is given by
    $\Gamma_\text{B}-\Gamma_\text{A}=S_\text{out,in,out}-S_\text{out,out,out}$,
    as indicated in the graph.}
    \label{F3}
\end{figure}

The third cumulant of the full-counting statistics, i.e. the first
correction term describing the deviation from the Gaussian
distribution of the charge $q$ transported through the conductor
during a time $t$, has been introduced in Refs.~\cite{levitov94}
and~\cite{levitov96}, and it is given by
\begin{equation} \label{cumulant1}
\begin{split}
\langle \langle q^3 \rangle \rangle = \int \limits_0^t dt_1 \int
\limits _0^t dt_2 \int \limits _0^t dt_3
   S_\text{K}^{(3)}(t_1,t_2,t_3),
\end{split}
\end{equation}
where the Keldysh-contour ordered correlation function is given by
\begin{equation} \label{Keldysh}
\begin{split}
S_\text{K}^{(3)}(t_1,t_2,t_3) &= \frac{1}{8}
   \langle
     \tilde{T}[\hat{I}_L(t_1)\hat{I}_L(t_2)\hat{I}_L(t_3)]
    +T[\hat{I}_L(t_1)\hat{I}_L(t_2)\hat{I}_L(t_3)]
  +3\tilde{T}[\hat{I}_L(t_1)\hat{I}_L(t_2)]\hat{I}_L(t_3)
  +3\hat{I}_L(t_1)T[\hat{I}_L(t_2)\hat{I}_L(t_3)]
   \rangle\\
&\quad
  -3\langle \hat{I}_L(t_1) \rangle
    \langle \hat{I}_L(t_2)\hat{I}_L(t_3) \rangle
  +2\langle\hat{I}_L(t_1)\rangle
     \langle\hat{I}_L(t_2)\rangle
      \langle\hat{I}_L(t_3)\rangle.
\end{split}
\end{equation}
Using the operator relations given by Eqs. (\ref{eq:two-current})
and (\ref{eq:three-current}), together with their anti-time-ordered
counterparts, and regrouping the current operators into deviation
operators $\delta\hat{I}_\text{in,out}(t) \equiv
\hat{I}_\text{in,out}(t)- \langle\hat{I}_\text{in,out}(t)\rangle$,
enables one to express this particular correlation function as
\begin{equation}
\begin{split}
S_\text{K}^{(3)}(t_1,t_2,t_3)
&=\langle
    \delta\hat{I}_{L,\text{in}}(t_1)
    \delta\hat{I}_{L,\text{in}}(t_2)
    \delta\hat{I}_{L,\text{in}}(t_3)
 -\frac{3}{4} \delta\hat{I}_{L,\text{in}}(t_1)
    \delta\hat{I}_{L,\text{in}}(t_2)
    \delta\hat{I}_{L,\text{out}}(t_3)
 -\frac{3}{2}\delta\hat{I}_{L,\text{in}}(t_1)
    \delta\hat{I}_{L,\text{out}}(t_2)
    \delta\hat{I}_{L,\text{in}}(t_3)\\
&\quad
 -\frac{3}{4} \delta\hat{I}_{L,\text{out}}(t_1)
    \delta\hat{I}_{L,\text{in}}(t_2)
    \delta\hat{I}_{L,\text{in}}(t_3)
 +\frac{3}{2}\delta\hat{I}_{L,\text{in}}(t_1)
    \delta\hat{I}_{L,\text{out}}(t_2)
    \delta\hat{I}_{L,\text{out}}(t_3)
 +\frac{3}{2}\delta\hat{I}_{L,\text{out}}(t_1)
    \delta\hat{I}_{L,\text{out}}(t_2)
    \delta\hat{I}_{L,\text{in}}(t_3)\\
&\quad
 -  \delta\hat{I}_{L,\text{out}}(t_1)
    \delta\hat{I}_{L,\text{out}}(t_2)
    \delta\hat{I}_{L,\text{out}}(t_3)
  \rangle.
\end{split}
\end{equation}
Each term here can now be expressed in terms of the Fourier transform
of the spectral function, Eq.~(\ref{eq:Ftrans}), and the time integrals
of Eq. \eqref{cumulant1} may be carried out explicitly. This results in
\begin{equation}
\begin{split}
&   \langle \langle q^3 \rangle \rangle
 = 2 \int \frac{d\omega_1}{2\pi}\int \frac{d\omega_2}{2\pi}
   S_\text{K}^{(3)}(\omega_1, \omega_2)
  \frac{\sin(\omega_2 t) - \sin(\omega_1 t)+ \sin[(\omega_1 - \omega_2) t]}
       {\omega_1 \omega _2 (\omega_1 -\omega_2)}
\end{split}
\label{eq:cum_int}
\end{equation}
where, for this particular ordering, we have
\begin{equation}
\begin{split}
S_\text{K}^{(3)}(\omega_1, \omega_2)
&=-\frac{3}{4} S_\text{in,in,out}(\omega_1, \omega_2)
  -\frac{3}{2} S_\text{in,out,in}(\omega_1, \omega_2)
  -\frac{3}{4} S_\text{out,in,in}(\omega_1, \omega_2)\\
&\quad
  +\frac{3}{2} S_\text{in,out,out}(\omega_1, \omega_2)
  +\frac{3}{2} S_\text{out,out,in}(\omega_1, \omega_2)
  - S_\text{out,out,out}(\omega_1, \omega_2).
\end{split}
\label{eq:S_K}
\end{equation}
This result is plotted in Fig.~\ref{F3}~(a) for energy-independent
scattering at zero temperature. Note that the multiplier of each
term in the sum above is obtained with the help of the binomial
distribution. The particular ordering for current operators, like
that in Eq. \eqref{Keldysh}, determines the final weight of each
$xyz$ spectral function.

\subsubsection{Asymptotic values of the third cumulant}

The third cumulant of FCS can be evaluated in the limits of both
short and long times $t$. For short $t$ the cumulant is determined
by the values of $S_\text{K}^{(3)}(\omega_1, \omega_2)$ at large
frequencies where $S_\text{out,out,out}(\omega_1, \omega_2)$ is
zero, and
\begin{equation}
\begin{split}
S_\text{K}^{(3)}(\omega_1, \omega_2)
 = 3 eV \frac{e^3}{h}\sum_n T_n(1-T_n)
 = 3 \Gamma_\text{A}
\end{split}
\end{equation}
nearly everywhere in the first quadrant of the
$(\omega_1,\omega_2)$--plane and zero elsewhere. Therefore, the
short-time value of the third cumulant is determined by the
$S_\text{in,out,out}$ and $S_\text{out,out,in}$ spectral functions
since only they have non-vanishing high-frequency values. We thus
have
\begin{equation}
\begin{split}
\langle \langle q^3 \rangle \rangle &\approx
   6 t eV \frac{e^3}{h}\sum_n T_n(1-T_n)
   \int_0^\infty \frac{\rmd x_1}{2\pi}\int_0^\infty \frac{\rmd x_2}{2\pi}
   \frac{\sin x_2 - \sin x_1 + \sin(x_1 - x_2)}
   {x_1 x _2 (x_1 -x_2)}\\
&= t eV \frac{e^3}{h} \sum_n T_n(1-T_n)
 = \Gamma_\text{A} t.
\end{split}
\end{equation}

Value of the third cumulant for large $t$ is obtained in a similar
manner. As long as $S_\text{K}^{(3)}(0,0)\ne 0$, the leading order
is given by
\begin{equation}
\begin{split}
&\langle \langle q^3 \rangle \rangle= 2 \int
\frac{d\omega_1}{2\pi}\int \frac{d\omega_2}{2\pi}
   S_\text{K}^{(3)}(\omega_1, \omega_2)
  \frac{\sin(\omega_2 t) - \sin(\omega_1 t)+ \sin[(\omega_1 - \omega_2) t]}
       {\omega_1 \omega _2 (\omega_1 -\omega_2)}\nonumber
\approx
   t S_\text{K}^{(3)}(0,0).
\end{split}
\end{equation}
For $k_BT \ll eV$, only $S_\text{out,out,out}(0,0)=-\Gamma_\text{B}$
has a non-vanishing value in $S_\text{K}^{(3)}$ at
$\omega_1=\omega_2=0$, and the linear growth at long times is then
given by
\begin{equation} \label{cum1}
\begin{split}
\langle \langle q^3 \rangle \rangle &= \Gamma_\text{B} t
\end{split}
\end{equation}
while, in  the opposite regime, $k_BT \gg eV$, the directed
three-current spectral functions become independent of the frequency
arguments, and $S_\text{K}^{(3)} = eV \frac{e^3}{h} \sum_n
T_n(1-T_n)$; the long term cumulant is then given by
\begin{equation} \label{cum2}
\langle \langle q^3 \rangle \rangle = \Gamma_\text{A} t.
\end{equation}
Since the Keldysh-ordered spectral function is independent of
frequency as long as $|\omega_1|,|\omega_2| \ll k_BT/\hbar$, this
result holds as long as $t\gg \hbar/k_BT$.

Both these results, Eqs. \eqref{cum1} and \eqref{cum2}, are in
agreement with those presented in Ref.~\cite{levitov04}, and thus
constitute a test of the correctness of our approach. Note in
particular that we find $\langle \langle q^3 \rangle \rangle/t =
\Gamma_\text{B} \sim \sum_n T_n(1-T_n)(1-2T_n)$ for low temperature.
This result has given rise to some discussion in the literature,
since Ref.~\cite{levitov92} obtained $\langle \langle q^3
\rangle \rangle/t \sim \sum_n T_n^2(1-T_n)$, different from
Eq.~\eqref{cum1}. Several authors~\cite{galaktionov03,lesovik03}
subsequently developed methods to analyze frequency-dependent
three-current correlation functions in order to assess the
correctness of Eq.~\eqref{cum1}. In Ref.~\cite{galaktionov03}
an effective action approach together with an involved
regularization procedure is used to establish Eq.~\eqref{cum1}.
According to Ref.~\cite{lesovik03} the frequency dependence of
$S_{\rm K}$, and hence the result for $\langle \langle q^3 \rangle
\rangle$, depends on the actual position of the spin-detector with
respect to the scatterer. Then, both results for $\langle \langle
q^3 \rangle \rangle$ cited above are found, depending on the
position of the detector. A drawback is that the specific
frequency-dependence of $S_{\rm K}$ postulated in
Ref.~\cite{lesovik03} generally does not conserve current. Let
us address the issue here in the framework of the in-out-ordering
technique. The $\sum_n T_n^2(1-T_n)$ proportionality is obtained in
Ref.~\cite{levitov92} by considering a straightforward
quantum analogue of the classical generating function, which leads
to the cumulant
\begin{equation}
\begin{split}
\langle\langle q^3 \rangle\rangle
= \left\langle \left( \int_0^t d\tau \delta\hat{I}(\tau) \right)^3
  \right\rangle
= \int_0^t dt_1 \int_0^t dt_2 \int_0^t dt_3
  \left\langle \delta\hat{I}(t_1) \delta\hat{I}(t_2) \delta\hat{I}(t_3)
  \right\rangle.
\end{split}
\label{qcub_un}
\end{equation}
Note that there is no specific time-ordering in this expression.
Use of
$\delta\hat{I}=\delta\hat{I}_\text{in}-\delta\hat{I}_\text{out}$
then leads to the entirely unordered correlation function
\begin{equation} \label{unordered}
\begin{split}
S_\text{Unordered}^{(3)}(t_1,t_2,t_3)
&=\langle
    \delta\hat{I}_{L,\text{in}}(t_1)
    \delta\hat{I}_{L,\text{in}}(t_2)
    \delta\hat{I}_{L,\text{in}}(t_3)
 -\delta\hat{I}_{L,\text{in}}(t_1)
    \delta\hat{I}_{L,\text{in}}(t_2)
    \delta\hat{I}_{L,\text{out}}(t_3)
 -\delta\hat{I}_{L,\text{in}}(t_1)
    \delta\hat{I}_{L,\text{out}}(t_2)
    \delta\hat{I}_{L,\text{in}}(t_3)\\
&\quad
 -\delta\hat{I}_{L,\text{out}}(t_1)
    \delta\hat{I}_{L,\text{in}}(t_2)
    \delta\hat{I}_{L,\text{in}}(t_3)
 +\delta\hat{I}_{L,\text{in}}(t_1)
    \delta\hat{I}_{L,\text{out}}(t_2)
    \delta\hat{I}_{L,\text{out}}(t_3)
 +\delta\hat{I}_{L,\text{out}}(t_1)
    \delta\hat{I}_{L,\text{out}}(t_2)
    \delta\hat{I}_{L,\text{in}}(t_3)\\
&\quad
 +\delta\hat{I}_{L,\text{out}}(t_1)
    \delta\hat{I}_{L,\text{in}}(t_2)
    \delta\hat{I}_{L,\text{out}}(t_3)
 -\delta\hat{I}_{L,\text{out}}(t_1)
    \delta\hat{I}_{L,\text{out}}(t_2)
    \delta\hat{I}_{L,\text{out}}(t_3)
  \rangle.
\end{split}
\end{equation}
The corresponding spectrum is given by
\begin{equation}
\begin{split}
S_\text{Unordered}^{(3)}(\omega_1, \omega_2)
&=-S_\text{in,in,out}(\omega_1, \omega_2)
  -S_\text{in,out,in}(\omega_1, \omega_2)
  -S_\text{out,in,in}(\omega_1, \omega_2)\\
&\quad
  +S_\text{in,out,out}(\omega_1, \omega_2)
  +S_\text{out,in,out}(\omega_1, \omega_2)
  +S_\text{out,out,in}(\omega_1, \omega_2)
  -S_\text{out,out,out}(\omega_1, \omega_2);
\end{split}
\label{eq:S_Un}
\end{equation}
it is plotted in Fig.~\ref{F3}~(b) for zero temperature. Here two
terms on the right hand side of~(\ref{eq:S_Un}) contribute at
zero frequency, namely $S_\text{out,in,out}$ and
$S_\text{out,out,out}$. For the unordered three current correlator,
we thus find that the corresponding third cumulant is given
asymptotically (for large $t$) by
\begin{equation}
\begin{split}
\langle \langle q^3 \rangle \rangle
&\approx
   (\Gamma_\text{B}-\Gamma_\text{A}) t
 =-t 2eV \frac{e^3}{h} \sum_n T_n^2(1-T_n),
\end{split}
\end{equation}
as found in Ref.~\cite{levitov92}. We therefore conclude
that the difference between this result and Eq.~\eqref{cum1} is
entirely due to the different ordering properties of the two
definitions of $\langle \langle q^3 \rangle \rangle$,
Eqs.~\eqref{qcub_un} and~\eqref{eq:cum_int}.

\subsubsection{Time-dependent third cumulant in various cases}

\begin{table*}
\begin{equation}
\begin{array}{lcccc}
   & \text{Tunnel junction,}
   & \text{Chaotic cavity}
   & \text{Diffusive wire} \\
   & 0\le T \ll 1
   & P(T)=\frac{1}{\pi}\frac{1}{\sqrt{T(1-T)}}
   & P(T)=\frac{l}{2L}\frac{1}{T\sqrt{(1-T)}}
\\\hline
\frac{1}{N}\langle \sum_n T_n \rangle
   & T_\text{ave}
   & \frac{1}{2}
   & \frac{l}{L}
\\
\frac{1}{N}\langle \sum_n T_n(1-T_n) \rangle
   & T_\text{ave}
   & \frac{1}{8}
   & \frac{1}{3}\frac{l}{L}
\\
\frac{1}{N}\langle \sum_n T_n(1-T_n)(1-2T_n) \rangle
   & T_\text{ave}
   & 0
   & \frac{1}{15}\frac{l}{L}
\\
\end{array}
\nonumber
\end{equation}
\caption{\label{tab:parameters} Values of the averaged transmission
parameters for three different types of noise sources: a tunnel
junction, a chaotic cavity, and a diffusive wire. Here $N$ is the
number of transmission channels, $P(T)$ is the distribution function
of transmission eigenvalues. In the case of a diffusive wire,
$L$ is the length of the wire and $l\ll L$ is the mean free path of
electrons.}
\end{table*}

We consider separately the time-dependent third cumulant generated
by three different kinds of noise sources: a tunnel junction, a
chaotic cavity and a diffusive wire~\cite{blanter00}, in the limit
where intrinsic dynamics and interaction effects can be ignored
(vanishingly small dwell and charge relaxation times) and scattering
can be considered as energy-independent. Then, the transmission
properties of these noise sources can be summarized as in Table
\ref{tab:parameters}.

\begin{figure}
\begin{center}
\includegraphics{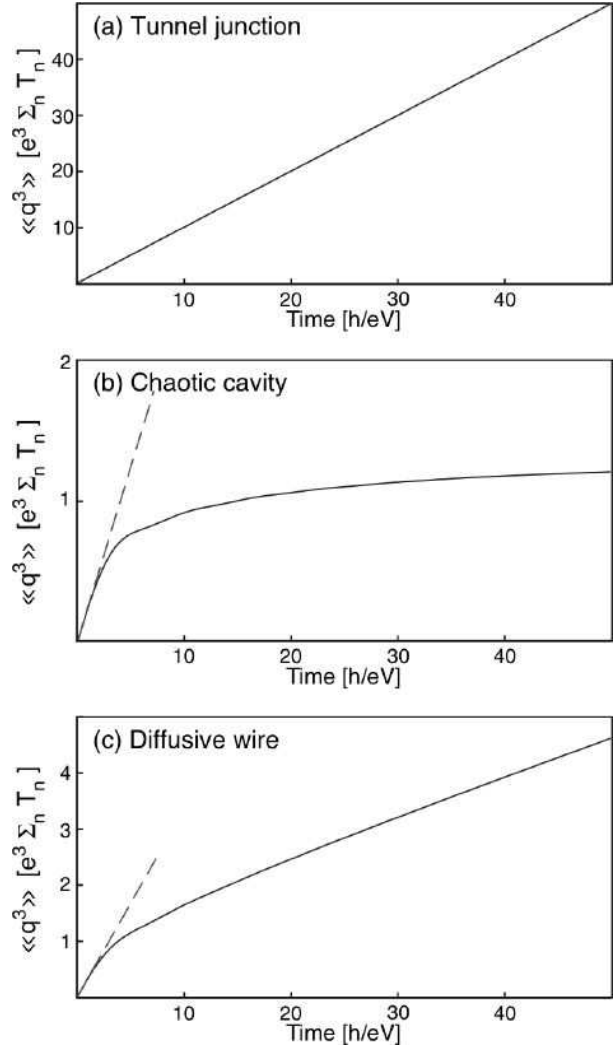}\\
\end{center}
\caption{\label{fig:cumulants}
  The third cumulant for (a) tunnel junction, (b) chaotic cavity and
  (c) diffusive wire at zero temperature. Both the tunnel junction and
  the diffusive wire show a linear growth at large $t$ due to
  non-vanishing zero-frequency value of $S_\text{out,out,out}$.}
\end{figure}

In an ideal {\em tunnel junction} all the transmission probabilities
are small, $T_n\ll 1$, and all the three relevant transmission
quantities are equal,
\begin{equation}
\sum_n T_n(1-T_n)(1-2T_n) \approx \sum_n T_n(1-T_n) \approx \sum_n
T_n \equiv N\,T_\text{ave}
\end{equation}
Here $N$ is the number of transport modes penetrating the tunnel
barrier. Hence, the linear coefficient of the time-dependent third
cumulant remains the same in both the small and long time limits.
Numerical integration of Eq. (\ref{eq:cum_int}) demonstrates only
this linear increase of the cumulant at all times, as illustrated in
Fig.~\ref{fig:cumulants}~(a).

As can be seen from Table~\ref{tab:parameters}, the transmission
probabilities of a {\em chaotic cavity} on the other hand are {\em
symmetrically} distributed between 0 and 1. Consequently, the
coefficient of the out-out-out noise term vanishes and the increase
of the third cumulant with time is slower than linear, see
Fig.~\ref{fig:cumulants}~(b).

Finally, for a {\em diffusive wire} the linear growth dominates
again for long times, after an initial transient up to several
$\hbar/eV$, as can be seen in Fig.~\ref{fig:cumulants}~(c).

\subsection{A fully symmetrized three-current correlation function}
A classical noise detector measures essentially a signal proportional
to the symmetrized two-current correlation function
\begin{equation} \label{symmcorr}
S_\text{Symm}^{(2)}(t_1,t_2)
= \frac{1}{2}
  \big\langle \hat{I}(t_1)\hat{I}(t_2)
              +\hat{I}(t_2)\hat{I}(t_1)
  \big\rangle
 -\big\langle \hat{I} \big\rangle^2.
\end{equation}
It is quite plausible to assume that a classical measurement of the
third-order correlations would yield a signal proportional to what
is essentially a generalization of~\eqref{symmcorr}, i.e., a fully
symmetrized three-current correlation function~\cite{golubev05}
\begin{equation}
\begin{split}
S_\text{Symm}^{(3)}(t_1,t_2,t_3) &= \sum_{i\ne j \ne k=1}^3
   \big\{ \frac{1}{16}\big\langle I(t_i)T[I(t_j)I(t_k)]\big\rangle
         +\frac{1}{16}\big\langle \tilde{T}[I(t_i)I(t_j)]I(t_k)\big\rangle\\
&\qquad\qquad\qquad
         -\frac{1}{2}\big\langle I(t_i)\big\rangle
                     \big\langle I(t_j)I(t_k)\big\rangle
         +\frac{1}{3}\big\langle I(t_i)\big\rangle
                     \big\langle I(t_j)\big\rangle
                     \big\langle I(t_k)\big\rangle
   \big\}.
\end{split}
\end{equation}
This correlator is indeed symmetric in all permutations of the time
arguments $t_1$, $t_2$, and $t_3$. We can then immediately rewrite the
corresponding spectral function with the help of the in-out~-ordering
technique as
\begin{equation}
\begin{split}
&  S_\text{Symm}^{(3)}(\omega_1,\omega_2)\\
&= \frac{1}{2} S_{\rm in,out,out}(\omega_1,\omega_2)
  +\frac{1}{2} S_{\rm in,out,out}(\omega_2-\omega_1,-\omega_1)\\
&\quad
  +\frac{1}{2} S_{\rm in,out,out}(-\omega_2,\omega_1-\omega_2)
  +\frac{1}{2} S_{\rm out,out,in}(-\omega_2,-\omega_1)\\
&\quad
  +\frac{1}{2} S_{\rm out,out,in}(\omega_1,\omega_1-\omega_2)
  +\frac{1}{2} S_{\rm out,out,in}(\omega_2-\omega_1,\omega_2)\\
&\quad
  -S_{\rm out,out,out}(\omega_1,\omega_2).
\end{split}
\label{eq:Zaikin}
\end{equation}
Here the presence of various combinations of $\omega_1$ and
$\omega_2$ is due to different orderings of the time arguments
$t_1$, $t_2$, and $t_3$, and they also give rise to the hexagonal
shape of the spectral function in the
$(\omega_1,\omega_2)$--plane. This result is plotted in
Fig.~\ref{F3}~(c) which coincides with the one found in
Ref.~\cite{golubev05}.

Comparing Eqs.~(\ref{eq:Zaikin}) and (\ref{eq:S_K}), or
Figs.~\ref{F3}~(c) and (a), we see that the symmetrized spectrum is
generally quite different from the Keldysh contour ordered one.
Nevertheless, the two coincide in the zero temperature, zero
frequency limit such that $S_\text{Symm}^{(3)}(0,0) = \Gamma_B$
and hence corresponds to the usual third cumulant of full counting
statistics.

\subsection{Three-current correlation functions of a multi-level quantum detector}
As it is well-known~\cite{aguado00,schoelkopf03}, two-level quantum
detector responds to two-current correlators such that the direct
transition rate to the higher level (absorption), given by the Fermi
golden rule, is normally determined by the non-symmetrized spectral
function
\begin{equation}
\begin{split}
S_\text{Q}^{(2)}(\omega)
= \int_{-\infty}^\infty dt
  e^{i\omega t} \langle \delta \hat{I}(t) \delta \hat{I}(0)\rangle
\end{split}
\end{equation}
at the frequency $\omega=-\Delta \omega$, where $\Delta \omega$ is
the level spacing. The corresponding relaxation rate (emission) is
given by the same spectral function but now at the frequency
$+\Delta \omega$. This result can be easily generalized to the case
of a multilevel detector.

The next-higher order correction to the transition rate, which
includes the effect of transitions via an intermediate state of a
multi-level detector, depends, among others, on the three-current
spectral function $S_\text{Q}^{(3)}(\omega_1,\omega_2)$, which was
recently discussed in~\cite{ojanen06,brosco06}
\begin{equation}
\begin{split}
S_\text{Q}^{(3)}(t_1,t_2,t_3)
= \frac{1}{(2\pi)^2} \int_{-\infty}^\infty d\omega_1 d\omega_2
  e^{-i\omega_1 (t_1-t_2)}e^{-i\omega_2 (t_2-t_3)}
  S_\text{Q}^{(3)}(\omega_1,\omega_2),
\end{split}
\end{equation}
where the partially time-ordered three-time current correlation function is
\begin{equation}
\begin{split}
S_\text{Q}^{(3)}(t_1,t_2,t_3)
&= \langle \delta I(t_1)T[\delta I(t_2)\delta I(t_3)]\rangle.
\end{split}
\end{equation}
We analyze this correlation function here using the in-out~-ordering
technique. Expanding in terms of in-out three current correlation functions yields
\begin{equation}
\begin{split}
S_\text{Q}^{(3)}(t_1,t_2,t_3)
&= \langle
    \delta I_\text{in}(t_1)\delta I_\text{in}(t_2)\delta I_\text{in}(t_3)
    -\delta I_\text{in}(t_1)\delta I_\text{out}(t_3)\delta I_\text{in}(t_2)
    -\delta I_\text{in}(t_1)\delta I_\text{out}(t_2)\delta I_\text{in}(t_3)\\
    &\qquad
    +\delta I_\text{in}(t_1)\delta I_\text{out}(t_2)\delta I_\text{out}(t_3)
    -\delta I_\text{out}(t_1)\delta I_\text{in}(t_2)\delta I_\text{in}(t_3)
    +\delta I_\text{out}(t_1)\delta I_\text{out}(t_3)\delta I_\text{in}(t_2)\\
    &\qquad
    +\delta I_\text{out}(t_1)\delta I_\text{out}(t_2)\delta I_\text{in}(t_3)
    -\delta I_\text{out}(t_1)\delta I_\text{out}(t_2)\delta I_\text{out}(t_3)
    \rangle
\end{split}
\end{equation}
such that the corresponding spectral function is
\begin{equation}
\begin{split}
S_\text{Q}^{(3)}(\omega_1,\omega_2)
&= S_\text{in,out,out}(\omega_1,\omega_2)
  +S_\text{out,out,in}(\omega_1,\omega_1-\omega_2)
  +S_\text{out,out,in}(\omega_1,\omega_2)
  -S_\text{out,out,out}(\omega_1,\omega_2),
\end{split}
\end{equation}
see Fig.~\ref{F3}~(d). The zero temperature,
zero frequency limit of this quantity is given by
$S_\text{Q}^{(3)}(0,0) = \Gamma_B$, i.e., it corresponds again to
the usual third cumulant of current statistics.

\subsection{Discussion}

Apart from the unordered spectral function Eq.~\eqref{eq:S_Un},
the various spectral functions discussed so far share many common
features at zero temperature: (i) None of them contains the
$S_\text{out,in,out}$ contribution. (ii) The sum of the terms
containing 0, 1, 2, and 3 out-currents are
given by binomial coefficients $(-1)^k \left(\begin{array}{c}3 \\
k\end{array}\right)$, where $k$ is the number of out-currents. For
energy-independent scattering, however, terms with $k=0,1$ vanish.
(iii) Regions for which $|\omega_{1,2}|\ge eV/\hbar$ are only
determined by the $k=2$ terms ($S_\text{in,out,out}$ and $S_\text{out,out,in}$)
while the zero-frequency value is given by the $k=3$ term
($S_\text{out,out,out}$). (iv) In
regions where $|\omega_{1,2}|\ge eV/\hbar$ the value of the spectral
function is either zero or it saturates to a constant, unlike the
two-current spectrum which increases linearly. The variously ordered
spectral functions differ mainly from each other based on how the
'spectral power' is distributed in the $(\omega_1,\omega_2)$--plane:
the quantum detector noise $S_\text{Q}^{(3)}$ has twice the value of
the symmetrized noise $S_\text{Symm}$, but that value is only
achieved for $\omega_1>0$ while the symmetrized noise has the
constant level everywhere in the $(\omega_1,\omega_2)$--plane,
except in the hexagonal area bound  within $|\omega_{1,2}|<
eV/\hbar$.

\section{Conclusions}

In this paper we have considered a formalism that facilitates
calculation of time-ordered current correlation functions and
applied it to current noise generated by a phase-coherent scatterer.
Causality of the real-time representation of the scattering matrix
causes products of in- and out-current operators,
$\hat{I}_\text{in}(t_1)$ and $\hat{I}_\text{out}(t_2)$, to vanish if
the in-current is taken later than the out-current; consequently,
time-ordering of current operators may be expressed using
in-out ordering, in which the out-current operators stand to the
left of the in-currents, and vice versa for anti-time-ordering. The
in-out ordering can be directly applied to current correlation
functions of arbitrary order, and they can be directly
evaluated in the case of thermal reservoirs. If the scattering
matrix is, furthermore, energy-independent the correlation functions
only depend on the transmission eigenvalues of the scatterer.

It is highly case-dependent to which particular current correlator a detector responds,
and we evaluate three alternative functions. While a {\it classical} noise detector would
respond to a fully symmetrized correlator, the spin detector discussed in the case of full
counting statistics depends on the Keldysh-contour-ordered correlation function and a
multi-level noise detector to a partially or fully time-ordered correlator.
We obtain all the answers without cumbersome time-ordered integrations.

\begin{acknowledgments}
We are indebted to M. B\"uttiker for pointing out
Ref.~\cite{beenakker01} to us as well as for fruitful
discussions that motivated us to carry out the work described in
this article.  We thank Academy of Finland for financial support.
F.W.J.H. acknowledges support from the EC-funded ULTI Project,
Transnational Access in Programme FP6 (Contract
\#RITA-CT-2003-505313) and from Institut Universitaire de France.
\end{acknowledgments}

\appendix
\section{\label{app:Wick} Calculation of the three-current
spectral functions with equilibrium reservoirs}

We follow Ref.~\cite{buttiker92}, and obtain all the three-current
spectral functions needed by applying Wick's theorem:
\begin{equation}
\begin{split}
& \left\langle
   \left(\hat{a}_k^\dagger \hat{a}_l - \langle \hat{a}_k^\dagger \hat{a}_l
   \rangle\right)
   \left(\hat{a}_m^\dagger \hat{a}_n - \langle \hat{a}_m^\dagger \hat{a}_n
   \rangle\right)
   \left(\hat{a}_p^\dagger \hat{a}_q - \langle \hat{a}_p^\dagger \hat{a}_q
   \rangle\right)
  \right\rangle\\
&= \langle \hat{a}_k^\dagger \hat{a}_q \rangle
   \langle \hat{a}_l \hat{a}_m^\dagger \rangle
   \langle \hat{a}_n \hat{a}_p^\dagger \rangle
  -\langle \hat{a}_k^\dagger \hat{a}_n \rangle
   \langle \hat{a}_l \hat{a}_p^\dagger \rangle
   \langle \hat{a}_m^\dagger \hat{a}_q \rangle\\
&= \delta_{kq}\delta_{lm}\delta_{np} f_k (1-f_m) (1-f_p)
  -\delta_{kn}\delta_{lp}\delta_{mq} f_k f_m (1-f_p).
\end{split}
\end{equation}
Next we insert this result into the expression of a three-current
correlation function, such as $\langle
\delta\hat{I}_{L,\text{in}}(t_1)
         \delta\hat{I}_{L,\text{in}}(t_2)
         \delta\hat{I}_{L,\text{in}}(t_3) \rangle$:
\begin{equation}
\begin{split}
&\langle \delta\hat{I}_{L,\text{in}}(t_1)
         \delta\hat{I}_{L,\text{in}}(t_2)
         \delta\hat{I}_{L,\text{in}}(t_3) \rangle\\
&= \frac{e}{h} \sum_l \int dE_1 dE_2 e^{i(E_1-E_2)t_1/\hbar}
   \frac{e}{h} \sum_n \int dE_3 dE_4 e^{i(E_3-E_4)t_2/\hbar}
   \frac{e}{h} \sum_p \int dE_5 dE_6 e^{i(E_5-E_6)t_3/\hbar}\\
&\qquad\times
    \bigg\langle
            \left( \hat{a}^{\dagger}_{L,l}(E_1)\hat a_{L,l}(E_2)
                  -\langle \hat{a}^{\dagger}_{L,l}(E_1)\hat a_{L,l}(E_2)\rangle \right)
            \left( \hat{a}^{\dagger}_{L,n}(E_3)\hat a_{L,n}(E_4)
                  -\langle \hat{a}^{\dagger}_{L,n}(E_3)\hat a_{L,n}(E_4)\rangle \right) \times\\
&\qquad\qquad\qquad
      \times\left( \hat{a}^{\dagger}_{L,p}(E_5)\hat a_{L,p}(E_6)
                  -\langle \hat{a}^{\dagger}_{L,p}(E_5)\hat{a}_{L,p}(E_6)\rangle\right) \bigg\rangle\\
&=-\left(\frac{e}{h}\right)^3 \sum_l \int dE_1 dE_3 dE_5
   e^{i(E_1-E_5)t_1/\hbar} e^{i(E_3-E_1)t_2/\hbar} e^{i(E_5-E_3)t_3/\hbar}
   f_{L}(E_1) f_{L}(E_3) (1-f_{L}(E_5))\\
&\quad
  +\left(\frac{e}{h}\right)^3 \sum_l \int dE_1 dE_3 dE_5
   e^{i(E_1-E_3)t_1/\hbar} e^{i(E_3-E_5)t_2/\hbar} e^{i(E_5-E_1)t_3/\hbar}
   f_{L}(E_1) (1-f_{L}(E_3)) (1-f_{L}(E_5))\\
&= \int \frac{d\omega_1}{2\pi} \frac{d\omega_2}{2\pi}
   e^{-i[\omega_1(t_1-t_2)+\omega_2(t_2-t_3)]}\\
&\qquad\qquad\times
   \frac{e^3}{h} N_L \int dE\
               f_{L}(E)
               [1-f_{L}(E+\hbar\omega_1)]
               [1-f_{L}(E+\hbar\omega_2)
                 -f_{L}(E+\hbar\omega_1-\hbar\omega_2)].\\
\end{split}
\end{equation}
from which we can infer that
\begin{equation}
\begin{split}
S_\text{in,in,in}(\omega_1,\omega_2)
&= \frac{e^3}{h} N_L \int dE\
               f_{L}(E)
               [1-f_{L}(E+\hbar\omega_1)]
               [1-f_{L}(E+\hbar\omega_2)
                 -f_{L}(E+\hbar\omega_1-\hbar\omega_2)],
\end{split}
\end{equation}
cf. Eq.~(\ref{eq:Ftrans}). We make next use of the following results valid for Fermi functions:
\begin{equation}
\begin{split}
& \int dE f(E)[1-f(E+\delta E_1)]
=  \frac{\delta E_1}{1-e^{-\beta \delta E_1}}
\\
& \int dE f(E)[1-f(E+\delta E_1)][1-f(E+\delta E_2)]
=  \frac{1}{1-e^{-\beta \delta E_1}}
   \left[ \frac{\delta E_2}{1-e^{-\beta \delta E_2}}
         -\frac{\delta E_2 -\delta E_1}
               {1-e^{-\beta (\delta E_2 - \delta E_1)}}\right]
\\
& \int dE f(E)[1-f(E+\delta E_1)]f(E+\delta E_2)
=  \frac{1}{1-e^{-\beta \delta E_1}}
    \bigg[ \delta E_1
          -\frac{\delta E_2}{1-e^{-\beta \delta E_2}}
          +\frac{\delta E_2 -\delta E_1}
           {1-e^{-\beta (\delta E_2 - \delta E_1)}}
    \bigg].\\
\end{split}
\end{equation}
The integration over energy $E$ can then be performed explicitly. In
this particular case of $S_\text{in,in,in}(\omega_1,\omega_2)$, the
energy integral contains Fermi functions of just one reservoir, and
therefore its value vanishes:
\begin{equation}
\begin{split}
\langle \delta\hat{I}_{L,\text{in}}(t_1)
        \delta\hat{I}_{L,\text{in}}(t_2)
        \delta\hat{I}_{L,\text{in}}(t_3) \rangle
= S_\text{in,in,in}(\omega_1,\omega_2)
= 0.
\end{split}
\end{equation}
This is generally true only for $S_\text{in,in,in}$ since it does
not depend on the possibly energy-dependent scattering matrix.
Spectral functions containing two in currents also have Fermi
functions of just the left reservoir, but the energy-dependence of
the scattering matrix may render the integrals non-zero. Yet in the
case of energy-independent scattering such spectral functions vanish.

\section{\label{app:spectral} In-out spectral functions of three currents}

In Table~\ref{tab:energydependent} all the eight different
three-current spectral functions are listed in the general case of
energy-dependent scattering and assuming equilibrium reservoirs. The
corresponding spectral functions for energy-independent scattering
are given in Table \ref{tab:energyindependent}, where $\{T_n\} $
denotes the set of energy-independent eigenvalues of the matrix
$t^\dagger t$.

\begin{table*}
\begin{equation}
\begin{array}{rl}
&S_\text{in,in,in}(\omega_1, \omega_2)=0\\
&S_\text{in,in,out}(\omega_1,\omega_2)\\
&= \frac{e^3}{h} \int \rmd E\
   \bigg\{\tr[s^{\dagger}_{LL}(E+\hbar \omega_2) s_{LL}(E)]
          f_L (E)[1-f_L(E+\hbar\omega_1)][1-f_L(E+\hbar\omega_2)]\\
&\qquad\qquad
  -\tr[s^{\dagger}_{LL}(E+\hbar \omega_1)
       s_{LL}(E+\hbar \omega_1-\hbar\omega_2 )]
   f_L(E)[1-f_L (E+\hbar \omega_1)]f_L (E+\hbar\omega_1-\hbar\omega_2)]
   \bigg\}\\
&S_\text{in,out,in}(\omega_1, \omega_2)\\
&= \frac{e^3}{h} \int \rmd E\
   \bigg\{\tr[s^{\dagger}_{LL}(E+\hbar \omega_1) s_{LL}(E+\hbar \omega_2)]
          f_L(E)[1-f_L (E+\hbar \omega_1)][1-f_L(E+\hbar \omega_2)]\\
&\qquad\qquad
  -\tr[s^{\dagger}_{LL}(E+\hbar \omega_1 - \hbar \omega_2) s_{LL}(E)]
   f_L(E)[1-f_L(E+\hbar\omega_1)]f_L(E+\hbar\omega_1-\hbar\omega_2)]
   \bigg\}\\
&S_\text{out,in,in}(\omega_1, \omega_2)\\
& = \frac{e^3}{h} \int \rmd E\
    \bigg\{\tr[s^{\dagger}_{LL}(E) s_{LL}(E+\hbar \omega_1)]
           f_L (E)[1-f_L (E+\hbar\omega_1)][1-f_L(E+\hbar \omega_2)]\\
&\qquad\qquad
   -\tr[s^{\dagger}_{LL}(E) s_{LL}(E+\hbar \omega_1)]
    f_L(E)[1-f_L(E+\hbar \omega_1)]f_L(E+\hbar\omega_1-\hbar \omega_2)]
    \bigg\}\\
&S_\text{in,out,out}(\omega_1, \omega_2)\\
&=\frac{e^3}{h} \sum_\alpha \int dE
  \bigg\{\tr[s^{\dagger}_{LL}(E+\hbar \omega_1)
             s_{L\alpha}(E+\hbar \omega_2)
             s^{\dagger}_{L\alpha}(E+ \hbar \omega_2)
             s_{LL}(E)]
         f_L (E)[1-f_L(E+\hbar \omega_1)][1-f_\alpha(E+\hbar \omega_2)]\\
&\qquad\qquad
        -\tr[s^{\dagger}_{LL}(E+\hbar \omega_1)
             s_{L\alpha}(E+\hbar \omega_1 -\hbar\omega_2 )
             s_{L\alpha}^\dagger(E+\hbar \omega_1 - \hbar \omega_2 )
             s_{LL}(E)]
         f_L(E)[1-f_L(E+\hbar\omega_1)]
         f_\alpha(E+\hbar\omega_1-\hbar\omega_2)]
  \bigg\}
\\
&S_\text{out,in,out}(\omega_1,\omega_2)\\
&= \frac{e^3}{h} \sum_{\alpha} \int\rmd E\
   \bigg\{\tr\left[s_{L\alpha}(E+\hbar\omega_1) s^\dagger_{\alpha L}(E)
                         s_{LL}(E+\hbar\omega_1) s^\dagger_{LL}(E)\right]
           f_\alpha(E) (1-f_L(E+\hbar\omega_1)) (1-f_L(E+\hbar\omega_2))\\
&\qquad\qquad
         -\tr\left[s_{LL}(E+\hbar\omega_1) s^\dagger_{LL}(E)
                   s_{L\beta}(E+\hbar\omega_1) s^\dagger_{\beta L}(E)\right]
                   f_L(E) (1-f_\beta(E+\hbar\omega_1))
                   f_L(E+\hbar\omega_1-\hbar\omega_2)
   \bigg\}\\
&S_\text{out,out,in}(\omega_1, \omega_2)\\
&= \frac{e^3}{h} \sum_\alpha \int dE
   \bigg\{\tr[s^{\dagger}_{LL}(E)
              s_{L\alpha}(E+\hbar\omega_1)
              s^{\dagger}_{L\alpha}(E+\hbar\omega_1)
              s_{LL}(E+\hbar \omega_2)]
          f_L (E)[1-f_\alpha(E+\hbar\omega_1)][1-f_L(E+\hbar \omega_2)]\\
&\qquad\qquad
         -\tr[s^{\dagger}_{L\alpha}(E)
              s_{LL}(E+\hbar\omega_1)
              s_{LL}^\dagger(E+\hbar\omega_1-\hbar\omega_2)
              s_{L\alpha}(E)]
          f_\alpha(E)[1-f_L(E+\hbar\omega_1)]
          f_L(E+\hbar\omega_1-\hbar\omega_2)]
  \bigg\}\\
&S_\text{out,out,out}(\omega_1, \omega_2)\\
&= \frac{e^3}{h} \sum_{\alpha\beta\gamma}\int \rmd E
   \bigg\{\tr[s^{\dagger}_{L\alpha}(E)
              s_{L \beta}(E+\hbar\omega_1)
              s^{\dagger}_{L\beta}(E+\hbar\omega_1)
              s_{L\gamma}(E+\hbar\omega_2)
              s_{L\gamma}^\dagger(E+\hbar\omega_2)
              s_{L\alpha}(E)]\\
&\qquad\qquad\qquad\qquad\times
          f_\alpha(E)[1-f_\beta(E+\hbar\omega_1)]
          [1-f_\gamma(E+\hbar\omega_2)]\\
&\qquad\qquad
         -\tr[s^{\dagger}_{L\alpha}(E)
              s_{L \beta}(E+\hbar\omega_1)
              s^{\dagger}_{L\beta}(E+\hbar\omega_1)
              s_{L\gamma}(E+\hbar\omega_1-\hbar\omega_2)
              s_{L\gamma}^\dagger(E+\hbar\omega_1-\hbar\omega_2)
              s_{L\alpha}(E)]\\
&\qquad\qquad\qquad\qquad\times
          f_\alpha(E)[1-f_\beta (E+\omega_1)]
          f_\gamma (E+\omega_1 - \omega_2)]
   \bigg\}.
\end{array}
\nonumber
\end{equation}
\caption{Three-current spectral functions for a general energy-dependent
scatterer.} \label{tab:energydependent}
\end{table*}

\begin{table*}
\begin{equation}
\begin{array}{ll}
S_\text{in,in,in}(\omega_1, \omega_2)
&= 0\\
S_\text{in,in,out}(\omega_1, \omega_2)
&= 0\\
S_\text{in,out,in}(\omega_1, \omega_2)
&= 0\\
S_\text{out,in,in}(\omega_1, \omega_2)
&= 0\\
S_\text{in,out,out}(\omega_1, \omega_2) &= \frac{e^3}{h} \sum_n
T_n(1-T_n) \int dE
   \bigg\{ f_L (E)[1-f_L(E+\hbar \omega_1)][1-f_R(E+\hbar \omega_2)]\\
&\qquad\qquad\qquad\qquad\qquad\qquad
          -f_L(E)[1-f_L (E+\hbar \omega_1)]f_R (E+\hbar \omega_1-\hbar\omega_2)]\bigg\}\\
S_\text{out,in,out}(\omega_1,\omega_2) &=\frac{e^3}{h} \sum_n
T_n(1-T_n) \int dE
   \bigg\{ f_{R}(E) [1-f_{L}(E+\hbar\omega_1)] [1-f_{L}(E+\hbar\omega_2)]\\
&\qquad\qquad\qquad\qquad\qquad\qquad
          -f_{L}(E)[1-f_{R}(E+\hbar\omega_1)] f_{L}(E+\hbar\omega_1-\hbar\omega_2)\bigg\}\\
S_\text{out,out,in}(\omega_1, \omega_2) &= \frac{e^3}{h} \sum_n
T_n(1-T_n) \int dE
   \bigg\{ f_L (E)[1-f_R(E+\hbar \omega_1)][1-f_L(E+\hbar \omega_2)]\\
&\qquad\qquad\qquad\qquad\qquad\qquad
          -f_R(E)[1-f_L (E+\hbar \omega_1)]f_L (E+\hbar\omega_1 - \hbar
          \omega_2)]\bigg\}\\
S_\text{out,out,out}(\omega_1, \omega_2) &= \frac{e^3}{h} \sum_n
T_n(1-T_n)^2 \int dE
   \bigg\{ f_L(E)
           [1-f_L (E+\hbar \omega_1)]
           [1-f_R (E+\hbar \omega_2)-f_R (E+\hbar\omega_1 - \hbar\omega_2)]\\
&\qquad\qquad\qquad\qquad\qquad\qquad\qquad
          +f_L(E)
           [1-f_R (E+\hbar \omega_1)]
           [1-f_L (E+\hbar \omega_2)-f_L (E+\hbar\omega_1 - \hbar\omega_2)]\\
&\qquad\qquad\qquad\qquad\qquad\qquad\qquad
          +f_R(E)
           [1-f_L (E+\hbar \omega_1)]
           [1-f_L (E+\hbar \omega_2)-f_L (E+\hbar\omega_1 - \hbar\omega_2)]
   \bigg\}\\
&\quad
   +\frac{e^3}{h} \sum_n T_n^2(1-T_n) \int dE
   \bigg\{ f_L(E)
           [1-f_R (E+\hbar \omega_1)]
           [1-f_R (E+\hbar \omega_2)-f_R (E+\hbar\omega_1 - \hbar\omega_2)]\\
&\qquad\qquad\qquad\qquad\qquad\qquad\qquad
          +f_R(E)
           [1-f_L (E+\hbar \omega_1)]
           [1-f_R (E+\hbar \omega_2)-f_R (E+\hbar\omega_1 - \hbar\omega_2)]\\
&\qquad\qquad\qquad\qquad\qquad\qquad\qquad
          +f_R(E)
           [1-f_R (E+\hbar \omega_1)]
           [1-f_L (E+\hbar \omega_2)-f_L (E+\hbar\omega_1 - \hbar\omega_2)]
\bigg\}
\end{array}
\nonumber
\end{equation}
\caption{Three-current spectral functions for an energy-independent
scatterer.} \label{tab:energyindependent}
\end{table*}

\end{document}